\documentstyle[12pt]{article}
\begin{document}
\null\vskip-24pt
\hfill SPIN-1999/17
\vskip-01pt
\hfill {\tt hep-th/9909197}
\vskip0.3truecm
\begin{center}
\vskip 2truecm
{\Large\bf
How to make the gravitational action on \\
\bigskip
non-compact space finite
}\\ 
\vskip 1.5truecm
{\large\bf
Sergey N.~Solodukhin\footnote{
email:{\tt S.Solodukhin@phys.uu.nl}}
}\\
\vskip 1truecm
{\it Spinoza Institute, University of Utrecht,\\
Leuvenlaan 4, 3584 CE Utrecht, The Netherlands}
\vskip 1truemm
\end{center}
\vskip 1truecm
\begin{abstract}
\noindent
The recently proposed technique to regularize the divergences of the gravitational action
on non-compact space by adding boundary counterterms is studied.
We propose prescription for constructing the boundary counterterms which are polynomial
in the boundary curvature. This prescription is efficient for both asymptotically Anti-de Sitter and 
asymptotically flat spaces. Being mostly interested in the asymptotically flat case
we demonstrate how our procedure works for known examples of non-compact spaces:
Eguchi-Hanson metric, Kerr-Newman metric, Taub-NUT and Taub-bolt metrics and others.
Analyzing the regularization procedure when boundary is not round sphere 
we observe that our counterterm helps to cancel large $r$ divergence of the action in the zero and first orders
in small deviations of the geometry of the boundary from that of the round sphere.
In order to cancel the divergence in the second order in deviations  a new
quadratic in boundary curvature counterterm is introduced. We 
argue that cancelation of the divergence for finite deviations possibly requires infinite series
of (higher order in the boundary curvature) boundary counterterms.
\end{abstract}
\begin{center}
{\it PACS number(s): 04.60.+n, 12.25.+e, 97.60.Lf, 11.10.Gh}
\end{center}
\vskip 1cm
\newpage
\baselineskip=.81cm
\section{Introduction}
\setcounter{equation}0
The classical dynamics of the gravitational field 
(metric $g_{\mu\nu}$ on $d$-dimensional manifold $M^d$)
is determined by the Einstein-Hilbert  (EH) action
\begin{equation}
W_{EH}[g]=-{1\over 16\pi G}\left( \int_{M^d}(R+2\Lambda )+2\int_{\partial {M^d}}K\right)~~,
\label{1}
\end{equation}
where the boundary term proportional to the extrinsic curvature $K$ of the boundary $\partial M$
should be added in order to  make the variational 
procedure of the action (when only metric but not its normal
derivative is fixed on the boundary) well defined \cite{GH}, \cite{Hawking}. 
When manifold $M$ is non-compact one considers a sequence of compact manifolds $M_r$
with boundary $\partial M_r$ parameterized by radius $r$ such that $M_r\rightarrow M$ for large $r$. 
The functional (\ref{1}) on non-compact manifold $M$ then should be understood as  result of the limit
$W_{EH}[M_r\rightarrow M]$. It is, however, a well known problem that this limiting procedure is not well
defined since $W_{EH}[M_r]$ diverges in the limit of large $r$. Therefore, the limiting procedure
should be accompanied with some regularization. 
The traditional way \cite{HH} of handling this problem is to subtract a contribution of 
some reference metric $g_0$ that matches suitably the asymptotic and topological properties
of the metric $g$. The choice of the metric $g_0$ is interpreted as fixing the vacuum state.
However, such a reference metric 
does not always exist
which makes this
subtraction procedure quite uncertain.

It was realized recently that when the space $M$ is asymptotically AdS (rather than asymptotically flat)
one can take an alternative route. 
In the context of AdS/CFT correspondence the general analysis (based on the previous mathematical works 
\cite{FG}, \cite{GrahamLee})
of the divergences of the EH action
for AdS space was done in \cite{HS}.
Inspired by AdS/CFT correspondence, Balasubramanian and Kraus \cite{BK}
have proposed to add to the action (\ref{1}) a counterterm which is functional of 
the curvature invariants of the induced metric $h_{ij}$ on $\partial M_r$. The role of this term
(which does not affect the gravitational equations in the bulk)
is to cancel appropriately the large $r$ divergence appearing in $W_{EH}[M_r]$. The counterterm 
$W_{ct}[h_{ij}]$ can be arranged as an expansion in powers of the curvature of the boundary metric. 
The first few terms are the following \cite{HS}, \cite{BK}, \cite{Myers}
\begin{equation}
W_{ct}^{bk}={1\over 16\pi G}\int_{\partial M^d_r}\sqrt{h}\left({2(d-2)\over l}+{l\over d-3}{\cal R}
+{l^3\over (d-5)(d-3)^2}({\cal R}_{ij}^2-{(d-1)\over 4(d-2)}{\cal R}^2)+...
\right)
\label{2}
\end{equation}
where ${\cal R}_{ij}$ and $\cal R$ are respectively Ricci tensor and   Ricci scalar of the boundary metric,
$l$ is the AdS radius related to the cosmological constant as $\Lambda ={(d-1)(d-2)\over 2l^2}$.
 The terms (\ref{2}) are sufficient
to cancel divergences for $d\leq 7$. On the other hand, the leading divergence in any $d$
is always killed by the term (first introduced in \cite{T}) 
in (\ref{2}) which is proportional 
to the area of the boundary\footnote{The extrinsic curvature of the asymptotic boundary of AdS space is constant,
$K={(d-1)\over l}$. Therefore, the first term in (\ref{2}) can be presented as 
surface integral of $K$. For $d=3$ this was observed in \cite{B}}.
It should be mentioned that introducing counterterms which are polynomial in the boundary curvature 
one is able to cancel all divergences of the action (\ref{1}) but not the logarithmic one
(appearing when $(d-1)$ is even). The later divergence can be canceled by adding a counterterm which is
not polynomial in curvature. For example, for $d=3$ it is the term ${\cal R}\ln {\cal R}$
that should be added. In higher dimensions there is ambiguity in choosing such terms. Up to this subtlety the 
procedure of introducing the counterterms (\ref{2}) is universal and well-defined.

Encouraged by this example one could try to construct appropriate  boundary term 
which cancels the leading 
divergence for asymptotically flat space. This term can be found but 
it is not analytic function of the boundary curvature \cite{Lau}, \cite{Mann}
\begin{equation}
W_{ct}^{LM}=-{c_{LM}\over 16\pi G}\int_{\partial M_r} \sqrt{{\cal R}}~~.
\label{3}
\end{equation}
The constant $c_{LM}$  depends on the topological type of the boundary at large $r$. 
For the Schwarzschild like  metric (when boundary is topologically $S_1\times S_{d-2}$)
one has $c_{LM}=-2\sqrt{{d-2\over d-3}}$.
Not requiring the counterterm to be analytic function of the curvature one can also construct a term 
interpolating between expressions (\ref{2}) and (\ref{3}) \cite{Mann}, \cite{KLS}
\begin{equation}
W_{ct}^{int}={1\over 16\pi G}~{2(d-2)\over l}\int_{\partial M_r} 
\sqrt{1+{l^2\over (d-3)(d-2)}{\cal R}+...}~~.
\label{4}
\end{equation}
Indeed, for large $r$ the boundary curvature $\cal R$ vanishes and we need to take the limit of small $\cal R$
in (\ref{4}) in order to get (\ref{2}). On the other hand, 
the asymptotically flat case is obtained by taking the
limit of large $l$ in (\ref{4}). The expression (\ref{3}) then is reproduced.
We stress that this interpolation exists only for the choice of the constant $c_{LM}$ in
(\ref{3}) as in the case of the Schwarzschild black hole. 
The boundary then is $S_1\times S_{d-2}$.
For other types of the boundary the expression (\ref{4}) 
does not match (\ref{3}) in the limit of large $l$.

There are, however, reasons to think that it is not eligible to drop the analyticity
in the proposed procedure of introducing 
the boundary counterterms. 
The form of the counterterms then is not unique and, in fact, quite ambiguous.
Indeed, for asymptotically flat space, not only
$\sqrt{\cal R}$ but any function $({\cal R}^2_{ij})^{1/4}$, $({\cal R}^2_{ijkl})^{1/4}$
or even higher roots of higher power curvature invariants can be chosen as a candidate
for the counterterm. In the asymptotically AdS case we also can take as a counterterm any
function $f(l^2{\cal R})$ that approaches  $(1+{l^2\over 2(d-3)(d-2)}{\cal R})$ for small $\cal R$.
Among these functions, in particular, there are ones which do not have the well-defined flat
space ($l\rightarrow \infty$) limit.

Another reason why it is not desirable to use non-analytic boundary counterterms appears from the
consideration of the EH term in quantum theory. Any quantum field makes contribution to the
EH action. In fact, this contribution is UV divergent and we have to renormalize the Newton's
constant $G$ (and cosmological term $\Lambda$) in order to handle these
divergences.  The natural question then if the structure of the classical action
$W_{EH}+W_{ct}$ is preserved under quantum corrections and whether it remains the same after
the renormalization. For the EH action (\ref{1}) along, this question was addressed in
\cite{BS}. It was found that the exact balance between the bulk and
boundary parts in the quantum action is the same as in the classical action (\ref{1}).
Hence the renormalization of only Newton's constant (the $\Lambda$ term was dropped in \cite{BS}
but this does not affect the main conclusion) is sufficient to regularize both the bulk and boundary 
UV divergences. In fact this statement is quite obvious in the case of
matter minimally coupled to gravity. One just has to impose Dirichlet or Neumann conditions
on the quantum field on the boundary $\partial M$. In the non-minimal case the boundary condition
should be chosen of the mixed type in order to make this statement valid.
Analyzing now this problem for the action $W_{EH}+W_{ct}$ with the counterterm in the form
(\ref{3}) or (\ref{4}) it is hard to imagine how this structure can be preserved
in the quantum case since only terms analytic in the boundary curvature are known to
appear in the quantum effective action (at least in its UV divergent part) 
on  manifold with boundary.

Concluding our brief analysis we see that the non-analytic boundary counterterms are likely not allowed
in unambiguous and universal procedure of the regularization of the gravitational action.
The purpose of this paper is to propose another way of constructing the counterterms
remaining in the class of the functions which are analytic in curvature. In asymptotically flat case,
we are not going to generalize the AdS prescription (\ref{2}) in the part of the dependence of the
boundary counterterm on the boundary metric. Instead, keeping
the general structure of the counterterm 
as in (\ref{2}), we define a scale parameter $l^*$ (analogous to the parameter $l$ in AdS case)
which characterizes the global geometry of the space (in fact, it is the  
coordinate invariant  $\it diameter$ of the space)
and can be used in the constructing the counterterms in the same fashion as 
in the AdS case. The prescription, thus, works universally both in asymptotically AdS
and asymptotically flat cases and deals only with analytic structure of the counterterms.

\section{The proposal}
\setcounter{equation}0
It should be noted that the counterterm (\ref{2}) is not an off-shell quantity. In fact, it 
contains some information about the asymptotic bulk geometry. Namely, the space-time 
is supposed to be Anti-de Sitter space with radius $l$. The role of the parameter $l$ in AdS space
is two-fold. First, it measures the curvature of the bulk geometry (Ricci scalar $R=-{d(d-1)\over l^2}$).
Second, it measures the size of the space: $l$ is that quantity which relates 
volume of AdS space $V(M_r)$
and area   of its boundary $A(\partial M_r)$ 
in the limit of large $r$, i.e. $l\sim {V(M_r)\over A(\partial M_r)}$.
This relation is the key one \cite{SW} in the  holographic correspondence
between the gravitational theory in the bulk of AdS space and the conformal field theory on the boundary.
As we said above, our idea is to introduce  parameter $l^*$
which for asymptotically flat space plays the role 
 similar to that of the parameter
$l$ in the AdS case. Since it is not possible in general to find any scale parameter universally
related to the curvature if the space is asymptotically flat it is the
holographic relation which we are going to generalize. Note, that in our prescription we do not
require metric to satisfy any equations of motion and in this sense it is off-shell
prescription. We only demand that (in the case of zero cosmological constant)
the curvature of the space-time dies sufficiently fast with large $r$ so that the bulk integral 
$\int_{M_r}R$ converges in the large $r$ limit. The only divergence of the gravitational action
(\ref{1}) then comes from the boundary term $2\int_{\partial M_r}K$.
Note also that we will be mostly considering    the leading divergence
of the action.

Consider compact manifold $M_r$ with boundary $\partial M_r$ parameterized by 'radial' coordinate
$r$ in an appropriate coordinate system.
Let $V(M_r)$ be invariant volume of $M_r$ and $A(\partial M_r)$ be area of the boundary $\partial M_r$.
Define the $\it diameter$ $l^*$ of the manifold $M_r$ as follows
\begin{equation}
l^*={V(M_r)\over A(\partial M_r)}~~.
\label{5}
\end{equation}
Consider now the sequence of compact manifolds $M_r$ approaching the noncompact manifold $M$
in the limit of large $r$. The diameter $l^*$ then, in general,  becomes 
function of $r$.
Defying the gravitational action $W_{gr}[M]$ as the limit of the actions $W_{gr}[M_r]$ 
for large $r$ we want it to be finite as $r\rightarrow \infty$. The  action
we propose takes the form
\begin{equation}
W_{gr}[M_r]=W_{EH}[M_r]+W_{ct}[\partial M_r]~~,
\label{6}
\end{equation}
where, as in the AdS case (\ref{2}), the boundary counterterm
\begin{equation}
W_{ct}[\partial M_r]=-{1\over 16\pi G} {c(\gamma )\over l^*_r}\int_{\partial M_r}\sqrt{h}
\label{7}
\end{equation}
is proportional to the area of the boundary.

First, we want to demonstrate that by adding the counterterm (\ref{7}) we do not change the Einstein
equations following from the EH action. We fix finite $r$ and consider small
variations of the metric in the bulk assuming the induced metric on the boundary $\partial M_r$
is fixed. The diameter (\ref{5}) changes under variation of the metric in the bulk.
At first sight it seems that this may result in rather complicated equations when the action (\ref{6})
is varied. However, it is quite surprising that the presence of the extra term (\ref{7}) makes the same affect in the field equations as that of the effective cosmological constant
$\Lambda_{eff}=-{c(\gamma )\over 2(l^*)^2}$:
\begin{equation}
\delta W_{gr}=-{1\over 16\pi G}\int_{M_r}\delta g^{\mu\nu}\left(R_{\mu\nu}-{1\over 2}g_{\mu\nu}R
-\Lambda_{eff}~g_{\mu\nu}\right)~~.
\label{var}
\end{equation}
So if we were considering the region of finite $r$ the extra boundary term (\ref{7})
would show up in the gravitational equations in the form of the finite cosmological
constant $\Lambda_{eff}$ (in fact, this would be an interesting mechanism of generating the
cosmological constant). However, being interested in the infinite space we should take the limit of
infinite $r$ (or infinite $l^*$ since $l^*\sim r$ 
for asymptotically flat space).
The quantity $\Lambda_{eff}$ disappears in this limit and the gravitational equations
remain unaffected.

In order to show that the gravitational action (\ref{6}), (\ref{7}) is indeed less divergent
than (\ref{1})
and determine the coefficient $c(\gamma )$ let us consider on $M^d$ the coordinate system
$(\chi , x^i, i=1,...,d-1)$ where  metric looks as
\begin{equation}
ds^2=d\chi^2+h_{ij}(\chi , x)dx^idx^j~~.
\label{8}
\end{equation}
The compact manifold $M_r$ is determined by range of
the radial coordinate $0\leq \chi \leq r$. The boundary $\partial M_r$ stays at
$\chi=r$ and $h_{ij}(r,x)$ is the metric induced on the boundary.

The area of the boundary $\partial M_r$ and the volume of $M_r$ are given by
\begin{eqnarray}
&&A(\partial M_r)\equiv A(r)=\int\sqrt{h(x,r)}d^{d-1}x ~~, \nonumber \\
&&V(M_r)\equiv V(r)=\int^rA(\chi )d\chi~~.
\label{9}
\end{eqnarray}
Assume that for large $r$ the area function $A(r)$ is represented by the series
\begin{equation}
A(r)=A_0r^\gamma+A_1r^{\gamma-1}+Br^{\gamma-1}\ln r+...~~,
\label{A}
\end{equation}
where $A_0,A_1, B$ are some coefficients and $...$ stands for the subleading terms.
Then for the volume of $M_r$ we have that
\begin{equation}
V(r)={A_0\over \gamma+1}r^{\gamma+1}+\left({A_1\over \gamma}-{B\over \gamma^2}\right)
r^{\gamma} + {B\over \gamma}\ln r+ ...
\label{V}
\end{equation}
The parameter $\gamma>0$ is the coordinate invariant, it shows how the area of $\partial M_r$  (or
volume of $M_r$) grows for large $r$.
The radius $l^*$ defined by the relation (\ref{5}) then reads
\begin{equation}
l^*={r\over \gamma+1}+\left({1\over \gamma (\gamma+1)}{A_1\over A_0}-{1\over \gamma^2}{B\over A_0}\right)
+{1\over \gamma (\gamma+1)}{B\over A_0}\ln r +...~~.
\label{10}
\end{equation}
For the extrinsic curvature of the boundary we have $\int_{\partial M_r}K=\partial_r A(r)$.
So that the leading divergence of the EH action
for large $r$ is proportional to $r^{\gamma-1}$.
Assuming that the bulk part of (\ref{6}) converges for large $r$ 
(this restricts the metric $h_{ij}(r,x)$ to grow asymptotically  not faster than $r^2$)
we find that the boundary part of (\ref{6}) is given by
\begin{equation}
W_{gr}^{boundary}=-{1\over 16\pi G}\left(2\partial_r A(r)+{c(\gamma )\over l^*(r)} A(r)\right)~~.
\label{B}
\end{equation}
Taking now the limit of infinite $r$ we find that
the leading divergence  of the gravitational action
cancels provided we choose the constant $c(\gamma )$ to be
\begin{equation}
c(\gamma )=-{2\gamma\over \gamma+1}~~.
\label{11}
\end{equation}
So that the regularized action 
\begin{equation}
W_{gr}={1\over 8\pi G}r^{\gamma-2}{1\over \gamma}B+
O(r^{\gamma-3})~~
\label{B1}
\end{equation}
is finite if $\gamma \leq 2$. In some cases the logarithmic term in the expansion (\ref{A}) is absent.
Then the leading term  in the action (\ref{B1}) is of the order $r^{\gamma-3}$. Thus, the
 adding the counterterm (\ref{7}) guarantees the cancelation of the leading divergence.
In order to kill
the still present in the action divergences 
one has to introduce extra counterterms
like the terms $l^*~\int_{\partial M}{\cal R}$ or $(l^*)^3\int_{\partial M}{\cal R}^2$. 
We consider such terms in sections 3 and 4.

In order to determine $l^*$ we have to
have information about the whole manifold $M$. However, for the cancelation of the divergences in the 
gravitational action only asymptotic behavior of $l^*$ is important. Therefore, 
it would be desirable to
define another quantity $l^*_a$
as asymptotic value of $l^*$ for large $r$. It can be used (instead of $l^*$) in constructing the
boundary counterterm (\ref{7}).
The advantage of using $l^*_a$ is that the counterterm (\ref{7}) then depends only on asymptotic 
properties of the bulk metric and is not sensitive to what happens inside the manifold.
The quantity $l^*_a$ is not, however, uniquely defined since it depends on how many 
terms (as in (\ref{10})) we want to keep
in the large $r$ expansion of $l^*$. On the other hand, the freedom in choosing
the coordinates in the asymptotic metric (\ref{8})
 also may result in  ambiguity in the definition of $l^*_a$. In all cases
these ambiguities affect only the subleading terms in $l^*_a$ and, eventually, in the
gravitational action. Note in this context that
picking up  the first three terms in the expansion (\ref{10})
and using this in the counterterm (\ref{7}) we get exactly the same result (\ref{B1})  for the
the leading part of gravitational action as when $l^*$ is used. 
Of course, we can add more subleading terms not changing this conclusion.

We still need to find an unambiguous and coordinate invariant notion of the asymptotic value
of $l^*$. In order to get an idea of such notion consider the large $r$
(i.e. valid outside of some compact large enough region of manifold $M$) expansion 
of the diameter $l^*$
done in any appropriate coordinate system. The $r$ dependent terms in this expansion,
for instance, functions $\{r, \ln r , 1, 1/r,1/r^2,... \}$, can be considered as forming 
basis in functional space and the large $r$ expansion of $l^*$ is just a decomposition of
$l^*$ along this basis. Among the elements consisting the basis
there are ones which grow infinitely with $r$. In the above example only the functions $r$ and $\ln r$ are such elements. Then, the projection  of $l^*$ 
onto subspace spanned by the asymptotically growing elements is what we will call the leading
asymptotic value  $l^*_a$. For the expansion (\ref{10}) we have
$$
l^*_{a}={r\over \gamma+1}
+{1\over \gamma (\gamma+1)}{B\over A_0}\ln r~~.
$$
Note, that by definition the constant term is not included in $l^*_a$. The quantity $l^*_a$
appears to be
unambiguous and coordinate invariant.

The gravitational action regularized by the counterterm (\ref{7}) with $l^*_{a}$ then reads
\begin{equation}
W^a_{gr}={1\over 8\pi G}(A_1-B) r^{\gamma-2}+
O(r^{\gamma-3})~~.
\label{B2}
\end{equation}
For $\gamma=2$ it takes the finite value which is different from  (\ref{B1}).
In many examples of $4$-dimensional  
metrics  we consider below the parameters $B$ and $A_1$ are related as 
$B=2A_1$. Then the limit of large $r$ in the  expressions (\ref{B1}) and (\ref{B2}) for $\gamma=2$
gives rise to results opposite in sign, $W_{gr}=-W_{gr}^a$. 
It happens that namely for $l^*_{a}$ our regularization
procedure gives the same result as  the standard subtraction method.
In all examples we present below the corresponding 
regularized action  is non-negative quantity.

As a simple illustration of our regularization procedure consider \\
the $d=4$ Schwarzschild metric
\begin{equation}
ds^2=g(\rho )d\tau^2+{d\rho^2\over g(\rho )}+\rho^2 ds^2_{S_2}~, ~~g(\rho )=1-{2m\over \rho }~~,
\label{*1}
\end{equation}
where $0\leq \tau\leq 2\pi\beta_H$, $\beta_H=4m$. It takes the form (\ref{8}) after performing 
the coordinate
transformation $\chi=\int^\rho {d\rho\over \sqrt{g(\rho)}}$. Asymptotically, we have
$\chi=\rho-m\ln\rho $.
The compact space $M_r$ is defined as $0\leq \chi\leq r$. The area of the boundary $\partial M_r$  
behaves as
\begin{equation}
A(r)=2\pi\beta_H \Sigma_2 (r^2-mr-2mr\ln r+...)
\label{*2}
\end{equation}
for large $r$. In (\ref{*2}) we 
recognize the expansion (\ref{A}) with $\gamma=2$. The defined above diameter $l^*$ and its
leading asymptotic values  are
\begin{eqnarray}
&&l^*_{}={r\over 3}-{m\over 3}\ln r +{m\over 3}+... ~~, \nonumber \\
&&l^*_{a}={r\over 3}-{m\over 3}\ln r ~~.
\label{*3}
\end{eqnarray}
Respectively, we have for the regularized action (\ref{B1}) and (\ref{B2})
\begin{equation}
W^{}_{gr}=-{4\pi m^2\over G}~,~~W^{a}_{gr}={4\pi m^2\over G}~~.
\label{*4}
\end{equation}
In $W^{a}_{gr}$ we recognize the standard expression for the action of the Schwarzschild
metric \cite{Hawking}, \cite{Hunter}, \cite{Mann}.

The parameter $\gamma$ in (\ref{A})-(\ref{10}) 
is an important characteristics of the asymptotic geometry
of the manifold $M^d$. 
Demanding the bulk metric to approach asymptotically the (locally) flat metric,
$\gamma$ is restricted by topology of the asymptotic
boundary $\partial M$.
In the simplest case, when size of $\partial M^d$ grows equally in all
directions (for example, if $\partial M^d$ is ($d-1$)-sphere and asymptotic metric on $M^d$ is 
$ds^2=dr^2+r^2ds^2_{S_{d-1}}$), we have $\gamma=d-1$. However, $\gamma$ can be 
less than $(d-1)$ if, for 
example, asymptotic metric is $ds^2=\sum_{i=1}^ndz_i^2+dr^2+r^2ds^2_{S_{d-n-1}}$, where each coordinate
$z_i$ is compact. Then $\gamma=d-n-1$. It seems to follow from these examples that $\gamma$
is related to the dimension of the spheric component
in the boundary $\partial M$. 
In $d=4$ the locally flat metric may take the form $ds^2=dr^2+r^2(d\theta^2+\sin^2 \theta d\phi^2)
+(d\tau+2n\cos\theta d\phi )^2$. The surface of constant $r$ is then the Hopf fiber bundle
$S_3\rightarrow S_2$ with fiber $S_1$. Locally it looks as direct product $S_1\times S_2$. However, the 
appropriate identifications (in $\tau$ and $\phi$) and overlapping coordinate patches give
the surface of constant $r$ the topology of $S_3$. In this case $\gamma=2$ is the same as for the
boundary $S_1\times S_2$.

In the asymptotically AdS case, the  expansion (\ref{A}) is not valid since the area
$$
A(r)=e^{(d-1)r/l}l^{d-1}A_0(1+O(e^{-r/l}))~~
$$
grows
exponentially with $r$.
As a result, the quantity $l^*$ asymptotically takes the constant value
\begin{equation}
l^*={l\over d-1}+O(e^{-r/l})~~.
\label{12}
\end{equation}
Therefore, the notion of the leading asymptotic value $l^*_a$ defined above is not good for
the asymptotically AdS spaces. In this case, we have to define it as the first, constant,
term in the large $r$ expansion. So that we have 
$l^*_a={l\over d-1}$. Using this quantity  
in the counterterm (\ref{7}) we reproduce correctly the first term in  the AdS expression (\ref{2})
provided value $\gamma=d-2$ is used to define the constant $c(\gamma )$ (\ref{11}).
It is interesting to note that using the quantity $l^*$ in (\ref{2})
we are able to cancel all divergences of the action including the logarithmic one.
In the  AdS case the size of the asymptotic boundary always grows in equal proportion in all
directions when boundary approaches infinity. So that $\gamma$ should depend only on 
dimension $d$ and be the same for all 
possible metrics on the boundary.
Note, that it is kind of mystery that $\gamma$ in the AdS case should be the same
as in the asymptotically flat case with one  $S_1$ component
in the boundary (i.e. as in the Schwarzschild black hole case). 
This becomes even more surprising when we recall that for the AdS space both the bulk and boundary
parts in the EH action diverge while for the asymptotically flat space only the boundary part causes
the divergence.
The same is also true for 
the Lau-Mann prescription with the counterterm (\ref{3}) where the coefficient $c_{LM}$ 
(for boundary being product of a sphere and $S_1$ factors)
should  in general be $c_{LM}=-2\sqrt{\gamma\over \gamma-1}$.
Only for $\gamma=d-2$ (the boundary is $S_1\times S_{d-2}$) 
there exists a correspondence between (\ref{3}) and the AdS prescription (\ref{4}).
There must be deep reasons 
for the coincidence of $\gamma$-s in these two cases.

\section{Examples}
\setcounter{equation}0

{\bf 3.1 Asymptotically (globally/locally) Euclidean spaces}

The asymptotically (globally) Euclidean space is defined  \cite{GP1} to be 
one admitting a chart $\{x^\mu\}$ such that for $(x_\mu x^\mu )^{1/2}=\rho>\rho_0$ the metric can be written as
\begin{equation}
g_{\mu\nu}=\left(1+{a^2\over \rho^2}\right)^2\delta_{\mu\nu}+O({1\over \rho^3})~~.
\label{3.1}
\end{equation}
It is known that the only asymptotically globally Euclidean solution of the Einstein equations
is flat space.  Usually the flat space-time is considered as a reference metric with respect to which one determines
the contribution of a curved metric to the action. In this way, one automatically (by definition)
assigns zero gravitational action to the flat space. It is then a  part of the Positive Action Theorem that
in the class of asymptotically Euclidean metrics the gravitational action is zero only if metric is flat.
In our method, however, flat space stands on the equal ground with any other space-times and it is not meaningless
to ask what is the gravitational action for the flat space-time itself. Choose metric on flat space $R^d$
to take the standard form $ds^2_{R^d}=d\chi^2+\chi^2 ds^2_{S_{d-1}}$ and determine the compact space 
$M_r^d$ as $0\leq \chi\leq r$. The space $M_r^d$ has volume
$V(r)={r^d\over d}\Sigma_{d-1}$ and its boundary $\partial M_r^d$ is round sphere $S_{d-1}$ with
area $A(r)=r^{d-1}\Sigma_{d-1}$
(we denote $\Sigma_n$ to be area 
of $n$-dimensional sphere, $\Sigma_3=2\pi^2$). 
So that the diameter (\ref{5}) of $M^d_r$ is $l^*={r\over d}$. 
We have $\gamma=d-1$ and $c(\gamma )=-{2(d-1)\over d}$ for
$M_r^d$. Substituting these ingredients into formula (\ref{B}) 
we find that the regularized gravitational action (\ref{6})
indeed vanishes for flat space.

In our analysis we are not restricted to consider  only solutions of the Einstein equations
and are interested in any metric for which the bulk integral $\int_{M_r}R$ converges
for large $r$.
An example of asymptotically Euclidean metric with vanishing Ricci scalar $R$ is the 
wormhole metric \cite{GP1}
\begin{equation}
ds^2=(1+{a^2\over 4\rho^2})^2\left(d\rho^2+\rho^2ds^2_{S_3}\right)~~,
\label{3.2}
\end{equation}
where $ds^2_{S_3}$ is metric of standard 3-sphere. Obviously, the condition (\ref{3.1}) is satisfied
for (\ref{3.2}). In fact, the metric (\ref{3.2}) describes space with two asymptotically Euclidean
regions at $\rho\rightarrow \infty $ and $\rho\rightarrow 0$ with minimal 3-sphere located at
$\rho=a/2$. One can bring metric (\ref{3.2}) to the form
(\ref{8}) by introducing radial coordinate $\chi=\rho-{a^2\over 4\rho}$. Then (\ref{3.2})
reads\footnote{Being extended to higher dimension $d$ the metric (\ref{3.3}) 
has scalar curvature $R={(d-1)(d-4)a^2\over (\chi^2+a^2)^2}$ and the integral 
$\int_{M_r}R$ diverges as $r^{d-4}$ for large $r$.}
\begin{equation}
ds^2=d\chi^2+(\chi^2+a^2)ds^2_{S_3}~~.
\label{3.3}
\end{equation}
Since the manifold has two asymptotic regions (at large negative and positive values of $\chi$)
we define the compact manifold $M_r$ in a symmetric way as 
$-r\leq \chi\leq r$. The boundary $\partial M_r$ then has two components at
$\chi=-r$ and $\chi=+r$ respectively. The manifold $M$ then is approached in the symmetric 
limit when $r\rightarrow \infty$.
The area $A(r)$ of the boundary $\partial M_r$ is $A(r)=2(r^2+a^2)^{3/2}\Sigma_3$.
The integral of the extrinsic curvature reads
$\int_{\partial M_r}K=\partial_r A(r)=6r(r^2+a^2)^{1/2}\Sigma_3$ and the
EH action $W_{EH}=-{3\over 4\pi G}r(r^2+a^2)^{1/2}\Sigma_3$ diverges as $r^2$ for large $r$.
Calculating the {\it diameter} $l^*$ (\ref{5}) of the manifold $M_r$ we find
\begin{equation}
l^*_r={r\over4}+{3\over 8}{a^2\over r}+O({1\over r^3})~~.
\label{3.4}
\end{equation}
In this case $\gamma=3$ and $c(\gamma )=-{3\over 2}$. It seems that our regularization
procedure applied to the metric (\ref{3.3}), according to (\ref{B1}),  should lead to 
the action which grows linear with  $r$. However,
for the metric (\ref{3.3}) the coefficients $A_1$ and $B$ in the expansion (\ref{A}) vanish.
Therefore, 
calculating  the regularized action (\ref{6})-(\ref{7}), (\ref{B}) we get
the finite value
\begin{equation}
W_{gr}=-{3\pi a^2\over 4G}~~
\label{3.5}
\end{equation}
when take the limit of infinite $r$. 
The leading asymptotic value for (\ref{3.4}) is $l^*_{a}={r\over 4}$. Using this quantity
in the boundary counterterm we find 
$$
W_{gr}^a={3\pi a^2 \over 4G}.
$$

If outside of a compact region metric approaches the standard flat $R^d$ metric with boundary $S_{d-1}$
identified under some
discrete subgroup of $SO(d)$ with a free action on $S_{d-1}$ 
such metric is asymptotically locally Euclidean. Note, that both for
the locally and globally Euclidean metrics  the parameter $\gamma$ in the large $r$ expansions
(\ref{A}), (\ref{V}) takes its maximal possible value $\gamma=d-1$.  
An example of $d=4$ asymptotically locally  Euclidean solution of the Einstein equations
is the Eguchi-Hanson metric \cite{GP1}, \cite{GH1}
\begin{eqnarray}
&&ds^2=(1-{a^4\over \rho^4})^{-1}d\rho^2+(1-{a^4\over \rho^4}){\rho^2\over 4}(d\psi+\cos\theta d\phi )^2
\nonumber \\
&&+{\rho^2\over 4}(d\theta^2+\sin^2\theta d\phi^2)~~,
\label{3.6}
\end{eqnarray}
where in order to remove the apparent singularity at $\rho=a$ one should identify $\psi$
modulo $2\pi$ rather than modulo $4\pi$ as is usual
for Euler angles on $S_3$. This identification makes the surface of constant $\rho>a$ into
projective space $RP^3$, i.e. 3-sphere with antipodal points identified. The 
surface $\rho=a$ is a 2-sphere.
Defying $M_r$ as $0\leq \rho\leq r$ we find
\begin{eqnarray}
&&V(r)={r^4\over 32}(1-{a^4\over r^4})\Sigma'_3~~, \nonumber \\
&&A(r)={r^3\over 8}(1-{a^4\over r^4})^{1/2}\Sigma'_3~~,
\label{3.7}
\end{eqnarray}
where $\Sigma'_3=\int\sin\theta d\theta d\phi d\psi$. From (\ref{3.7}) the diameter $l^*$ is found to be
$l^*_r={r\over 4}(1-{a^4\over r^4})^{1/2}$. Vector normal to $\partial M_r$ has the components
$(n^r=(1-{a^4\over r^4})^{1/2},0,0,0)$ and we have that $\int_{\partial M_r}K=n^r\partial_r A(r)=
{3\over 8}r^2-{1\over 8}{a^4\over r^2}+O({1\over r^6})$. Thus, the EH action for the metric (\ref{3.6})
diverges as $r^2$. Calculating the regularized action (\ref{B}) (in this case $\gamma=3$) we find that the 
counterterm (\ref{7}) precisely compensates the $r^2$-divergence while the rest terms vanish (as ${a^4\over r^2}$) in the 
limit of large $r$. Thus, the metric (\ref{3.6}) has vanishing gravitational action,
$W_{gr}=0$. One obtains the same result if the asymptotic quantity $l^*_{a}$ is used in the counterterm.

\bigskip

{\bf 3.2 Asymptotically flat spaces} 

In the class of asymptotically flat metrics we include all metrics describing  space-time  with
boundary  at infinity being an $S_1$ bundle over an $S_{d-2}$, where $S_1$ fiber approaches a constant length.
So the growth of the area of the boundary for large $r$ is due to the spheric component $S_{d-2}$
and we have $\gamma=d-2$ for all spaces of this class. For $d=4$  such
bundles are labeled by the first Chern number. If it vanishes the boundary has
topology of the direct product $S_1\times S_{d-2}$. Otherwise, its topology is more complicated.
The boundary then is  a squashed sphere. The fiber $S_1$ in the bundle is usually due to the compactified 
Euclidean time.

{\bf A. Schwarzschild metric in $d$ dimensions}

A generalization of the 4-dimensional metric (\ref{*1}) to higher dimensions is the metric
\begin{equation}
ds^2=g(\rho )d\tau^2+{d\rho^2\over g(\rho )}+\rho^2 ds^2_{S_{d-2}}~, ~~g(\rho )=1-({\mu\over \rho })^{d-3}~~,
\label{3.8}
\end{equation}
where $0\leq \tau\leq 2\pi\beta_H$, $\beta_H={2\mu\over d-3}$.
Though the analysis can be done in terms of the metric of the type (\ref{8}) the calculation is simpler
for the metric in the form (\ref{3.8}).
In this coordinate system we define the compact manifold $M_r$
 as $\mu\leq \rho \leq r$. The area of $\partial M_r$ and volume of $M_r$ are
\begin{eqnarray}
&&A(r)=2\pi\beta_H \Sigma_{d-2} r^{d-2}g^{1/2}(r)~~, \nonumber \\
&&V(r)=2\pi\beta_H\Sigma_{d-2} {1\over d-1}(r^{d-1}-\mu^{d-1})~~.
\label{3.9}
\end{eqnarray}
For large $r$ the area $A(r)$ grows as $r^{d-2}$ so that\footnote{Since $\gamma=d-2$ it seems that the 
regularized action (\ref{B1}) should diverge as $r^{\gamma-2}$. However, it happens that 
for the metric (\ref{3.8}) the only non-zero (growing with $r$) terms in
the expansion (\ref{A}) for the area are 
$r^{\gamma-2}$ and  $r$.
 Therefore, the action is indeed finite.} 
$\gamma =d-2$.
The diameter $l^*$ of $M_r$ is
\begin{equation}
l^*(r)={r\over d-1}g^{-1/2}(r)\left(1-({\mu\over r})^{d-1}\right)~~.
\label{3.10}
\end{equation}
In the coordinate system (\ref{3.8}) the integral of the extrinsic curvature of the 
boundary is given by the formula
$\int_{\partial M_r}K=n^r\partial_r A(r)$, where $n^r=g^{1/2}(r)$ is the non-zero component 
of vector  normal to $\partial M_r$.
For finite $r$ the regularized  action (\ref{6}) reads
$$
W_{gr}=-{\beta_H \Sigma_{d-2}\over 8G}\left((d-3)\mu^{d-3}-{2(d-2)\over r^2}\mu^{d-1}{g(r)\over 1-({\mu\over r})^{d-1}}\right)~~.
$$
In the limit of large $r$ it goes to the finite value 
\begin{equation}
W_{gr}=-{(d-3)\mu^{d-3}\over 8G}\beta_H \Sigma_{d-2}~~.
\label{3.11}
\end{equation}
The asymptotic value of the diameter  (\ref{3.10}) is 
$
l^*_{a}={r\over d-1}~~.
$
It can be used in the constructing the boundary counterterm (\ref{7}). The corresponding regularized action
$$
W_{gr}^a=-{\beta_H \Sigma_{d-2}\over 8G}\left( (d-3)\mu^{d-3}+2(d-2)r^{d-3}(g(r)-g^{1/2}(r))
\right)~~
$$
exactly coincides with the one obtained   within the standard subtraction procedure
$W=-{1\over 8\pi G}\int_{\partial M_r}(K-K_0)$ provided the reference metric is 
metric of flat space with $K_0={d-2\over r}$.
For large $r$ we obtain
\begin{equation}
W^a_{gr}={\mu^{d-3}\over 8G}\beta_H \Sigma_{d-2}~~.
\label{3.12}
\end{equation}

\bigskip

{\bf B. Taub-NUT and Taub-bolt metrics} 

For $d=4$ the boundary at infinity, which is  the fiber bundle of $S_1$ over $S_2$, may be non-trivial if the corresponding Chern number is non-zero. It is the case for the Taub-NUT and Taub-bolt metrics which can be 
present in the form \cite{GH1}
\begin{equation}
ds^2=f(\rho )(d\tau+2n\cos \theta d\phi )^2+{d\rho^2\over f(\rho )}+(\rho^2-n^2)ds^2_{S_2}~~,
\label{Taub}
\end{equation}
where the metric function is
\begin{equation}
f(\rho )={\rho-n\over \rho+n}
\label{f1}
\end{equation}
for Taub-NUT metric and
\begin{equation}
f(\rho )={(\rho-n/2)(\rho-2n)\over (\rho^2-n^2)}
\label{f2}
\end{equation}
for Taub-bolt. The Euclidean time $\tau$ in (\ref{Taub}) should be identified with period $8\pi n$ while
the angle $\phi$ is identified modulo $2\pi$. In fact one should take two
different coordinate patches which are non-singular at northpole ($\theta=0$) and southpole
($\theta=\pi$) respectively. The overlapping these patches gives the surface of constant $\rho>\rho_+$
($\rho_+=n$ for Taub-NUT and $\rho_+=2n$ for Taub-bolt) the topology of $3$-sphere
(with $(\tau/(2n),\theta ,\phi )$ being Euler angles).

The manifold $M_r$ in the sequence of spaces approaching the space $M$ is defined  by the range 
$\rho_+\leq \rho\leq r$ of the radial coordinate.
The square root of determinant of metric (\ref{Taub})
$\sqrt{g}=(r^2-n^2)\sin\theta$
does not depend on the   metric function $f(r)$.  Therefore both for (\ref{f1}) and (\ref{f2})
the volume of the space $M_r$ is
\begin{equation}
V(r)=32\pi^2n\left(({r^3\over 3}-{\rho^3_+\over 3})-n^2(r-\rho_+)\right)~~.
\label{VT}
\end{equation}
The area of the boundary $\partial M_r$ is
\begin{equation}
A(r)=32\pi^2n (r^2-n^2)f^{1/2}(r)~~.
\label{AT}
\end{equation}
The diameter  of $M_r$ then is
$$
l^*_{TN}={1\over 3}(r+2n)\sqrt{r-n\over r+n}= {r\over 3}+{n\over 3}-{n^2\over 2r}+O(1/r^2)
$$
for the Taub-NUT space and
$$
l^*_{TB}={r\over 3}+{5\over 12}n-{7\over 32} {n^2\over r}+O(1/r^2)
$$
for Taub-bolt. In both cases the leading asymptotic value is $l^*_a={r\over 3}$.
Calculating the regularized gravitational action one
obtains
\begin{equation}
W_{gr}^a=-W_{gr}={4\pi n^2\over G}
\label{W1}
\end{equation}
for the Taub-NUT metric and 
\begin{equation}
W^a_{gr}=-W_{gr}={5\pi n^2\over G}
\label{W2}
\end{equation}
for the Taub-bolt metric. The expressions for $W^a_{gr}$ agree
with the results obtained in \cite{Mann} 
within the square-root prescription (\ref{3})
and with the calculation performed in \cite{Myers} using the AdS prescription.
In the later case the expressions (\ref{W1}) and (\ref{W2}) are recovered in the limit of infinite
AdS radius $l$. 
The difference between (\ref{W2}) and (\ref{W1}) yields the results of \cite{Hunter}, 
\cite{Hawking-Hunter}.
On the other hand, (\ref{W1}), (\ref{W2}) agree with
the much older result by Gibbons and Perry \cite{GP} obtained by an ``imperfect match''
of the Taub-NUT solution to Euclidean flat space.

\bigskip

{\bf C. Kerr-Newman metric} 

The Euclidean Kerr-Newman metric parameterized by mass $m$, electric charge $q$ and the rotation parameter
$a$ takes the form
\begin{eqnarray}
&&ds^2=g_{rr}dr^2+g_{\theta\theta}d\theta^2 + g_{\tau \tau}d\tau^2+2g_{t\phi}dtd\phi +g_{\phi\phi}d\phi^2
\nonumber \\
&&g_{rr}={\rho^2 \over \Delta}~~,~~ g_{\theta\theta}=\rho^2~~,~~g_{\tau\tau}=\rho^{-2}( \Delta+
a^2\sin^2\theta ) ~~; \nonumber \\
&&g_{\tau\phi}= \rho^{-2} \sin^2\theta (r^2-a^2-\Delta )~, ~~g_{\phi\phi}=\rho^{-2}\left(
(r^2-a^2)^2+\Delta a^2\sin^2 \theta\right) \sin^2 \theta~, \nonumber \\
&&\Delta (r)=r^2-a^2-q^2-2mr~, ~~\rho^2=r^2-a^2 \cos^2 \theta~~.
\label{Kerr}
\end{eqnarray}
This metric has vanishing scalar curvature although the Ricci tensor vanishes only if
$q=0$.
The Euclidean  metric (\ref{Kerr}) can be obtained from the Lorentzian
metric by taking the Wick rotation of the time  and supplementing this by
the parameter transformation $a\rightarrow \imath a$, $q\rightarrow \imath q$.
The Euclidean instanton (\ref{Kerr}) is regular manifold for $r\geq r_+$, where
$r_{+}=m+ \sqrt{m^2+a^2+q^2}$ is the positive root of equation $\Delta ( r)=0$, provided one made certain
identifications. The angle coordinate $\phi$ should be identified modulo
$2\pi$. We must also identify points $(\tau , \phi )$ with $(\tau+2\pi \beta_H,\phi +2\pi \Omega
\beta_H )$, where
$$
\beta_H={(r^2_+-a^2)\over \sqrt{m^2+a^2+q^2}}~,~~ \Omega={a\over r^2_+-a^2}~~.
$$
For the metric (\ref{Kerr}) one has that $\sqrt{g}=\rho^2\sin\theta$.
The boundary of the manifold $M_r$
we define by equation $r=const$. We then find for the volume and area
\begin{eqnarray}
&&V(r)={8\over 3}\pi^2\beta_H \left(r^3-r^3_+-a^2(r-r_+)\right)~~, \nonumber \\
&&A(r)=4\pi^2\beta_h\Delta^{1/2}(r)I(r)~~,
\label{VA}
\end{eqnarray}
where $I(r)=\int_{-1}^{1}dx\sqrt{r^2-a^2x^2}$. From (\ref{VA}) we find
$$
l^*(r)={r\over 3}+{m\over 3}+({m^2\over 2}+{q^2\over 6}-{a^2\over 9}){1\over r}+O(1/r^2)
$$
for the diameter of $M_r$. The asymptotic value is $l^*_a={r\over 3}$.

Vector normal to the boundary $\partial M_r$ has component $n^r=({\Delta\over \rho^2})^{1/2}$ so the 
extrinsic curvature of the boundary $K=\nabla_\mu n^\mu={1\over \sqrt{g}}\partial_r(\sqrt{g}n^r)$.
For the  metric $h_{ij}$ induced on $\partial M_r$ we have 
$\sqrt{h}=\sqrt{g}({\Delta\over \rho^2})^{1/2}$ and hence, 
$K\sqrt{h}={r\Delta (r)\over \rho^2}\sin\theta +{1\over 2}\Delta'\sin\theta$.
Performing the integration over angle $\theta$ and taking into account that
$\int d\phi d\tau=4\pi^2\beta_H$ we obtain
$$
\int_{\partial M_r}K\sqrt{h}=4\pi^2\beta_H \left( {\Delta (r)\over a}\ln ({r+a\over r-a})+\Delta'(r)\right)~~.
$$
Calculating now the regularized action we get
\begin{equation}
W^a_{gr}=-W_{gr}=\pi m\beta_H~~.
\label{KW}
\end{equation}
This reproduces the previous results \cite{Hawking}, \cite{Hunter} obtained within
the subtraction regularization procedure.

\bigskip

{\bf D. Flat space in spheroidal coordinates} 

In all examples   considered so far the sequence of boundaries $\partial M_r$
was chosen in a natural way for given  metric on manifold $M$.
However, there, of course, exists freedom
to choose different shape for boundaries $\partial M_r$.
The important question arises as how the limiting value of the
regularized action $W_{gr}[M_r\rightarrow M]$ depends on the limiting
sequence of the boundaries chosen. To address this question we consider
flat $d$-dimensional space. In the Section 3.1 we demonstrated that choosing
$\partial M_r$ to be the round
$(d-1)$-sphere of radius $r$ the limiting value of the gravitational action is just zero.
In this section we want to re-calculate the action  for flat space 
choosing the set of boundaries $\partial M_r$ to be now sequence of  spheroidal surface.
We choose the  spheroidal  coordinates
on flat space where metric reads
\begin{eqnarray}
&& ds^2={\rho^2\over \Delta (r)}dr^2+\rho^2d\theta^2+\Delta (r)\sin^2\theta d\phi^2
+r^2\cos^2\theta ds^2_{S_{d-3}}~~,
\label{SF}
\end{eqnarray}
where 
$\Delta (r)=r^2-a^2$, $\rho^2=r^2-a^2\cos^2\theta$. The metric (\ref{SF})
is obtained as $\tau=const$ part of the  $(d+1)$-dimensional metric \cite{MP} generalizing the Kerr metric  
by setting mass to zero. It is regular if $r\geq a$. For (\ref{SF}) we
have $\sqrt{g}=\rho^2 r^{d-3} \sin\theta \cos^{d-3}\theta$.
We define the surface $\partial M_r$ by equation\footnote{The surface of constant $r$
is spheroidal surface with curvature depending on angle $\theta$. For $d=3$ we have, in particular,
${\cal R}={2r^2\over (r^2-a^2\cos^2\theta )^2}$. Note that the in the limit of infinite $r$ the surface tends to
the round sphere.}
 $r=const$. The further calculation goes along the same
lines as for the Kerr-Newman metric. We have for the volume and area respectively
\begin{eqnarray}
&&V(r)={4\pi\Sigma_{d-3}\over d(d-2)}  r^{d}\left(1-{a^{2}\over r^2}\right)
~~, \nonumber \\
&&A(r)=4\pi\Sigma_{d-3} {r^{d-1}\over (d-2)}
\left(1-({d-1\over d}){a^2\over r^2}-{1\over d(d+2)}{a^4\over r^4}+O(a^6/r^6)\right)~~.
\label{SVA}
\end{eqnarray}
The large $r$ expansion for the diameter $l^*$
is
\begin{equation}
l^*(r)={r\over d}\left(1-{1\over d}{a^2\over r^2}-{(d^2-2)\over d^2(d+2)}{a^4\over r^4}+O(a^6/r^6)\right)~~.
\label{lva}
\end{equation} 
Hence, its asymptotic value is $l^*_a={r\over d}$ as it should be for $\gamma=d-1$.
For the extrinsic curvature of $\partial M_r$ we have
$K\sqrt{h}=n^r\partial_r(\sqrt{g}n^r)$, where $n^r=({\Delta \over \rho^2})^{1/2}$
is component of normal vector. After performing the  integration over $\tau, \phi $ and $\theta$ we get
$$
\int_{\partial M_r} K\sqrt{h}=4\pi \Sigma_{d-3} r^{d-2}\left({d-1\over d-2}-{(d-1)\over d}{a^2\over r^2}-{2\over d(d+2)}{a^4\over r^4}+O(a^6/r^6)
\right)~~.
$$
Computing in the leading order  the regularized action 
\begin{eqnarray}
&&W^a_{gr}=-{\Sigma_{d-3}\over 2G}{(d-1)\over d(d-2)}a^2r^{d-4}(1+O(a^2/r^2))~~, \nonumber \\
&&W_{gr}={ \Sigma_{d-3}\over 2G} {(d+1)\over d^2(d+2)}a^4r^{d-6}(1+O(a^2/r^2))~~
\label{ws}
\end{eqnarray}
we find that the result is different for $W_{gr}$ and $W^a_{gr}$. The action $W_{gr}$ is less singular:
the leading term is $r^{d-6}$. Let us concentrate on analysis of the regularized action $W_{gr}$.
We observe that
for $d\leq 5$
the limiting value of the action is zero. 
This is in agreement with our  computation done in Section 3.1 for the boundary being
the round sphere.
For $d=6$ the large $r$ behavior of the action is dominated
by the constant term and the limiting value is finite.
For $d\geq 7$ the gravitational action  diverges as $r^{d-6}$.
These observations are similar to the ones made  in \cite{KLS}.
We see that for $d\geq 6$
the  limiting value of the action 
appears to  depend on the choice of the limiting sequence of boundaries.
A possible resolution of this problem is
to find a new counterterm which may kill the leading term in (\ref{ws}).
It is not difficult to find the appropriate counterterm 
among terms quadratic in the boundary curvature. Indeed, the invariant
$S\equiv ({\cal R}_{ij}^2-{1\over d-1}{\cal R}^2)$
identically vanishes for the round $(d-1)$-dimensional sphere
(in this case ${\cal R}_{ij}={(d-2)\over r^2}\gamma_{ij}$). 
Moreover, its first variation $\delta S$ due to small deviation of the metric
from that of the round sphere also identically vanishes. Therefore, the geometric invariant $S$
is non-zero only in the second order with respect to the deformation of the metric of round sphere.
The spheroidal surface
is  $a^2$-dependent deformation of the round sphere and
we can expect that in the leading order (small $a$ or large $r$) 
the invariant $S$ is proportional to  $a^4$. This is exactly what we need for the cancelation of
the divergence (\ref{ws}) also proportional to $a^4$.
The detail computation shows that for large $r$ we have
$$
S={(d-3)^2\over (d-1)}\left((d-2)\cos(\theta)^4-2\cos(\theta)^2+(d-2)\right){a^4\over r^8}
+O({a^6\over r^{10}})~~.
$$
Integrating this expression over angles we find that the boundary integral
\begin{equation}
W_{ct1}=-{1\over 16\pi G}{d^2(d-1)\over (d-3)^2(d-2)}(l^*)^3
\int_{\partial M_r}\left({\cal R}_{ij}^2-{1\over d-1}{\cal R}^2
\right)
\label{ct1}
\end{equation}
is that additional counterterm which can be added to the gravitational action
 in order to cancel the divergence (\ref{ws}).
Some similarity of the counterterm (\ref{ct1}) and the quadratic in the boundary curvature counterterm
in the AdS prescription (\ref{2}) should be noted.

\bigskip

\section{More of boundary geometry}
\setcounter{equation}0

{\bf 4.1 Asymptotic geometry of Ricci flat space} 

The universality of the AdS prescription (\ref{2}) valid for any metric
$h_{ij}(x)$ on the asymptotic boundary of AdS space is based on the possibility to 
well pose the Dirichlet boundary problem for the Einstein equations with positive
cosmological constant. Indeed, the solution of the Einstein equations is completely
determined once one fixes the induced metric on the boundary of the space
(more precisely, one needs to find the manifold of negative constant curvature which has a given
conformal structure at infinity). An existence theorem for such Einstein metric was
proved in \cite{GrahamLee}. One can explicitly obtain an asymptotic expansion of bulk metric near infinity starting from any metric at infinity \cite{HS}. Technically, one uses the distinguished coordinate system 
\cite{FG} where the bulk metric takes the form
\begin{equation}
ds^2={l^2d\rho^2\over 4\rho^2}+{1\over \rho}\left(h_{ij}(x)+h^{(1)}_{ij}(x)\rho +h^{(2)}_{ij}(x)\rho^2+...\right)dx^idx^j~~,
\label{4.1}
\end{equation}
$\rho$ is radial coordinate, $\rho=0$ determines the infinity of the space. 
Once one picks the first term $h_{ij}(x)$ the Einstein equations determine the other terms in the $\rho$-expansion,
$h^{(1)}_{ij}(x),h^{(2)}_{ij}(x),...$, as local covariant functions of the metric $h_{ij}(x)$.
That is why the divergences (due to the integration in the action over small $\rho$)
of the EH action are completely determined by the asymptotic metric $h_{ij}(x)$ and expressed as local
covariant functions of $h_{ij}(x)$. The idea of introducing the counterterms
determined on the regularized (staying at $\rho=\epsilon$) boundary (with boundary metric
${1\over \epsilon}h_{ij}(x)$) then appears quite naturally \cite{BK}.

To what extent the same can be done in asymptotically flat case? The Einstein equations then
determine metric with vanishing Ricci tensor. In analogy with Anti-de Sitter case
we can fix metric $h_{ij}(x)$ at infinity of the space and try to determine the  metric in the bulk by solving
the equation $R_{\mu\nu}=0$. In particular, we can use coordinate system $(\rho ,x)$  where the metric
takes the form similar to (\ref{4.1})
\begin{equation}
ds^2=d\rho^2 +\rho^2\left(h_{ij}(x)+h^{(1)}_{ij}(x){1\over \rho}+h^{(2)}_{ij}(x){1\over \rho^2}+...\right)
dx^idx^j~~,
\label{4.2}
\end{equation}
the infinity of the space is at infinite value of $\rho$.
[Note, that the boundary area for the metric (\ref{4.2}) grows as $\rho^{d-1}$ so that 
the metric is characterized by
value $\gamma=d-1$ 
of the parameter $\gamma$ introduced in Section 2.]
We do not give here the detail analysis of the problem. We just note that the bulk equations $R_{\mu\nu}=0$
determine not only the $\rho$-evolution of the metric (i.e. how terms in the expansion (\ref{4.2}) are 
determined by the asymptotic metric $h_{ij}(x)$) but also give constraint on the
asymptotic metric $h_{ij}(x)$. Indeed, the first term in the series (\ref{4.2}) can not be
arbitrary: the bulk equations $R_{\mu\nu}=0$ dictate it to satisfy the equation
\begin{equation}
{\cal R}_{ij}[h]=(d-2)h_{ij}~~.
\label{4.3}
\end{equation}
Thus,  infinity of Ricci flat space should
have geometry of $(d-1)$-dimensional de Sitter space which we will loosely  call round 
sphere. It  is  an important difference from the Anti-de Sitter case where the
boundary metric can be arbitrary from given conformal class.
We may also consider the case when  the boundary has topology of product of $n$ circles $S_1$ (radius of each circle
approaches constant value at infinity) and $(d-n-1)$-dimensional
surface $\Sigma$. The parameter $\gamma=d-n-1$ in this case. The bulk metric then is Ricci-flat only if
surface $\Sigma$ is $(d-n-1)$-dimensional de Sitter space.
 
The constraint (\ref{4.3}) explains why for known examples our prescription (\ref{7})
gives the same result as the Lau-Mann prescription (\ref{3}). In these cases 
the boundary is chosen consistently with the form of the metric (\ref{4.2}),
i.e. it is defined as $\rho=r=const$ (the boundary metric is $r^2h_{ij}(x)$). From  
(\ref{4.3}) we get that the integral in (\ref{3}) is
proportional to the area of the boundary in the same way  as in 
the prescription (\ref{7}).

\bigskip

{\bf 4.2  When boundary is not round sphere}

We saw above that infinity of  the Ricci flat space has
geometry of sphere $S^{d-1}$. Therefore, it is distinguished choice to take 
boundary $\partial M_r$ to be a round sphere. 
This is what we were doing in the most examples considered above.
However, we are of course free to
choose boundary to be any other closed surface $\Sigma$, topologically equivalent to sphere.
An important question then is how much our regularization procedure is sensitive to this.
In particular, it was claimed in Section 2 that the coefficient $c(\gamma )$ in front of the
counterterm (\ref{7}) is determined only by topology but not geometry of the boundary.
However, the analysis in Section 2 was done for the boundary
chosen consistently with the form of the bulk metric (\ref{8}), i.e. defined as 
$\chi=const$. In Ricci flat case, as we just have seen, it  means that the  boundary is round sphere.
So, what happens if we take arbitrary surface as the boundary?
Should the coefficient in front of the counterterm (\ref{7})  depend on geometry of the surface?
All these questions, in fact, challenge the universality of our prescription.
A related important question (partially addressed in section 3) 
concerns the gravitational action of flat space: does it universally vanish
or it may take non-zero or even infinite value depending on the shape of the boundary.

Analyzing this problem we consider the simplest possible case when manifold $M^d$ is flat space with Cartesian
coordinates $\{z_1,z_2,...,z_d\}$. In this space we consider the sequence of surfaces $\Sigma_r$ defined by 
equation\footnote{I would like to thank Rob Myers for suggestion to consider this example
and for many comments on the subject of this section.} 
\begin{equation}
{z^2_1\over a_1^2}+{z_2^2\over a^2_2}+...+{z^2_d\over a^2_d}=r^2~~.
\label{4.4}
\end{equation}
and parameterized by radius $r$, when $r$ goes to infinity the region $M_r$
inside the surface (\ref{4.4}) covers the whole manifold $M$. The parameters
$\{a_i\}$ indicate how the surface (\ref{4.4}) deviates from the round sphere $S^{d-1}$
which is determined by equation (\ref{4.4}) when all $a_i=1, i=1,..,d$. 
It should be noted that the sequence of surfaces (\ref{4.4}) is quite different from the spheroidal
surfaces considered in section 3. The spheroidal surfaces still approach the round sphere when radius
goes to infinity while the surface (\ref{4.4}) remains different from the round sphere even at 
infinite $r$.

The Einstein-Hilbert action
for the region $M^d_r$ is given by
\begin{equation}
W_{EH}=-{1\over 16\pi G}\int_{\Sigma_r}2K~~.
\label{4.4*}
\end{equation}
The trace of the extrinsic curvature $K$ of the surface (\ref{4.4}) is
\begin{equation}
K={(d-2+\sum_i {1\over a^2_i})\over (\sum_i{z^2_i\over a^4_i})^{1/2}}-
{(\sum_i{z^2_i\over a^6_i})\over (\sum_i {z^2_i\over a^4_i})^{3/2}}~~.
\label{4.5}
\end{equation}
For the integral of
(\ref{4.5}) over the surface (\ref{4.4}) we get the expression
\begin{equation}
\int_{\Sigma_r}K\sqrt{\gamma}=(d-1)r^{d-2}\Sigma_{d-1}\alpha_d(a_i)~~,
\label{4.6}
\end{equation}
where $\alpha_d(a_i)$ is some function of parameters $\{ a_i \}$. The integration in
(\ref{4.6}) can be performed after introducing on surface (\ref{4.4}) appropriate system of
angle coordinates and reduces to elliptic type integrals. When $d=4$ and only one parameter $a_1=a$
in (\ref{4.4}) differs from $1$  the integration can be done explicitly and we get
$$
\alpha_4(a)={2\over 3}(a+{1\over a+1})~~.
$$
It seems that in order to cancel the divergence (\ref{4.6}) with help of counterterm
(\ref{7}) we need to assume that the coefficient in front of the integral in (\ref{7})
explicitly depends on parameters $\{ a_i \}$. This would indicate that our regularization prescription (\ref{7})
is not universal and applies (as it stands) only to boundary which is round sphere while 
in more general case
the prescription should be modified appropriately to the concrete geometry of the boundary.

It is, however, instructive to analyze the behavior of the EH action (\ref{4.4*})-(\ref{4.5})
and the regularized action (\ref{6})-(\ref{7}) (with $c(\gamma)=d-2$ as it stands for the round sphere)\
with respect to small deviations of parameters $\{a_i\}$ from $1$. For simplicity, let us assume
that only two parameters $a_1=a,a_2=b$ are different from $1$. Then up to second order in
$(a-1)$ and $(b-1)$ we find that
\begin{eqnarray}
&&\alpha_d(a,b)=1+{(d-2)\over d}[(a-1)+(b-1)) 
+{(d-2)\over d(d+2)}((a-1)^2+(b-1)^2 \nonumber \\
&&+{(d^2-d-8)\over (d-1)}(a-1)(b-1)]+O((a-1),(b-1))^3
\label{4.7}
\end{eqnarray}
for the function $\alpha_d(a_i)$ appearing in (\ref{4.6}). On the other hand,
we have 
\begin{equation}
V={ab\over d}r^d\Sigma_{d-1}
\label{4.8}
\end{equation}
for the volume of $M_r^d$ and
\begin{eqnarray}
&&A=r^{d-1}\Sigma_{d-1} [1+{(d-1)\over d}((a-1)^2+(b-1))+{(d-1)\over 2d(d+2)}((a-1)^2+(b-1)^2
\nonumber \\
&&+2({d^2-5\over d-1})(a-1)(b-1)+O((a-1),(b-1))^3 ]
\label{4.9}
\end{eqnarray}
for the area of the surface $\Sigma_r$. We are now in a position
to compute the regularized gravitational action 
(\ref{6})-(\ref{7}) 
$$
W_{EH}+W_{ct}=-{1\over 16\pi G}\left(\int_\Sigma 2K-{2(d-1)\over d~l^*}A\right)~~.
$$
We find that
\begin{equation}
W_{EH}+W_{ct}=-{4(d^2-1)\over d^2(d+2)}\left((a-1)^2+(b-1)^2-{2\over (d-1)}(a-1)(b-1)\right)r^{d-2}\Sigma_{d-1}
~~.
\label{4.10}
\end{equation}
We see that (\ref{4.10}) is still divergent but it is now quadratic in $(a-1)$ and $(b-1)$.
So, the counterterm (\ref{7}) cancels the large $r$ divergence of (\ref{4.4*}) in the
zero and  first orders in $(a-1)$ and $(b-1)$!
Is it possible to find a counterterm which may cancel
the divergence (\ref{4.10}) in the second order in $(a-1)$ and $(b-1)$? The answer is yes.
The required counterterm is exactly the term (\ref{ct1}) which we introduced earlier
in order to cancel the divergences for the spheroidal boundary.
Even   the overall (dependent on dimension $d$) coefficient in (\ref{ct1}) takes the right form. 
In order to
prove the last statement we present here the result of integration of the invariant
$S\equiv {\cal R}^2_{ij}-{1\over d-1}{\cal R}^2$ over the surface (\ref{4.4}). In the second order in $(a-1)$
and $(b-1)$ it reads
\begin{equation}
\int_{\Sigma_r}S={4(d-3)^2(d-2)(d+1)\over d(d+2)}r^{d-5}\Sigma_{d-1}
\left((a-1)^2+(b-1)^2-{2\over (d-1)}(a-1)(b-1)\right)~~.
\label{4.11}
\end{equation}
We see that the divergence of (\ref{4.10}) is exactly canceled in the second order  by the counterterm (\ref{ct1})
so that in the functional
$$
W_{EH}+W_{ct}+W_{ct1}=r^{d-1}\Sigma_{d-1}~O((a-1),(b-1))^3
$$
the divergence may appear only in the third and higher orders in $(a-1)$ and $(b-1)$. In fact this is true
in general  when all parameters $\{a_i\}$ are different from $1$. In this case the expressions
(\ref{4.10}) and (\ref{4.11}) are proportional to the same  symmetric combination
$[\sum_i (a_i-1)^2-{1\over (d-1)}\sum_{i\neq j}(a_i-1)(a_j-1)]$ and 
the cancelation of the divergences is evident.

Presumably, this can be continued further: we should introduce more (higher order in the boundary
curvature) counterterms in order to cancel the large $r$ divergence in next orders
in $(a_i-1)$. What we get then is infinite series of the counterterms so that
the large $r$ divergence of the action cancels in all orders in $(a_i-1)$. 
This would mean that once the gravitational action includes the whole infinite series of the counterterms
it vanishes for flat space for any choice of the (topologically equivalent to sphere)
boundary.
We expect that structure of this infinite 
series is universal and determined only by topology of the boundary. It would be interesting to
get more terms in this series and  see if it converges to some compact expression
of the boundary curvature.

\section{Acknowledgments}

I would like to thank Andrei Barvinskii and Kostas Skenderis for useful discussions
and Robert Mann for reading the manuscript and
valuable comments. I am especially grateful to Rob Myers for reading the
manuscript, many interesting comments and encouragement.

\end{document}